\documentclass[]{cJAS2e}

\usepackage{subfigure}
\usepackage{times,epsfig,makeidx,amsfonts,amsmath,dcolumn,enumerate,multirow,graphicx,amssymb,colortbl,bm,url}
\usepackage{mathtools,upgreek}
\usepackage{afterpage,lscape}
\usepackage{paralist}
\usepackage{xr}

\theoremstyle{plain}
\newtheorem{theorem}{Theorem}

\newtheorem{corollary}{Corollary}

\theoremstyle{remark}

\begin{document}



\title{{\itshape Flexible objective Bayesian linear regression with applications in survival analysis}}

\author{Francisco J. Rubio$^{\rm a}$$^{\ast}$\thanks{$^\ast$Corresponding author. Email: Francisco.Rubio@lshtm.ac.uk
\vspace{6pt}} and Keming Yu$^{\rm b}$\\\vspace{6pt}  $^{a}${\em{London School of Hygiene \& Tropical Medicine, WC1E 7HT, UK}};\\
$^{b}${\em{ Department of Mathematical Sciences, Brunel University, Uxbridge UB8 3PH, UK.}}\\\received{v4.1 released December 2013} }

\maketitle

\begin{abstract}
We study objective Bayesian inference for linear regression models with residual errors distributed according to the class of two-piece scale mixtures of normal distributions. These models allow for capturing departures from the usual assumption of normality of the errors in terms of heavy tails, asymmetry, and certain types of heteroscedasticity. We propose a general noninformative, scale-invariant, prior structure and provide sufficient conditions for the propriety of the posterior distribution of the model parameters, which cover cases when the response variables are censored. These results allow us to apply the proposed models in the context of survival analysis. This paper represents an extension to the Bayesian framework of the models proposed in \cite{RH15}. We present a simulation study that shows good frequentist properties of the posterior credible intervals as well as point estimators associated to the proposed priors. We illustrate the performance of these models with real data in the context of survival analysis of cancer patients.
\begin{keywords}
Accelerated failure time model; residual life; noninformative prior; predictive; two-piece distributions.
\end{keywords}

\begin{classcode}\textit{Classification codes}: 62E15; 62N01; 62N02; 62P10
\end{classcode}

\end{abstract}

\section{Introduction}\label{intro}

The use of normal residual errors in linear regression models (LRMs) is perhaps the most common distributional assumption. However, the normality assumption can be inappropriate in practice given that the inference about the regression parameters is affected when the true distribution of the errors is asymmetric or heavy tailed. In order to overcome this shortcoming, alternative distributional assumptions have been proposed. We refer the reader to \cite{RH15} for an extensive review of the different distributional assumptions which include, for instance, the family of scale mixtures of normals (SMN) \citep{W84,FS00,VS15}, skew-elliptical and skew-symmetric distributions \cite{SDB03,AG08,A08,RG16}, semiparametric approaches such as quantile regression \cite{YM01,L12,RS13}, among others.

In a Bayesian framework, it is often of interest to employ noninformative priors; for instance, when the prior knowledge about the model parameters is vague. These kinds of priors are functions of the parameters, not necessarily integrable, that induce a well-defined posterior distribution with good frequentist properties. In this direction, \cite{FS00} proposed an improper prior structure for LRMs with residual errors distributed according to the family of SMN. In the context of survival regression models, \cite{VS15} studied the use of Jeffreys-type priors for accelerated failure time (AFT) models (which are LRMs
for the logarithm of a set of survival times) with SMN errors. However, the use of noninformative priors in LRMs with flexible errors that allow for capturing skewness has received little attention. In this line, \cite{RD14} proposed an improper prior structure for AFT models with errors distributed according to the generalised extreme value distribution. They provided a list of sufficient conditions for the propriety of the corresponding posterior distribution which involves truncating the parameter space. Recently, \cite{RG16} proposed a general noninformative prior structure for LRMs with skew-symmetric errors. They provided conditions for the propriety of the posterior distribution that cover cases where the response variables are censored.

In this paper, we study the use of the class of two-piece scale mixtures of normal (TPSMN) distributions for modelling the residual errors in LRMs from a Bayesian perspective. These sorts of distributional assumptions enjoy several advantages. First, this family of error models contains the class of SMN distributions as a particular case, which has been used to account for the presence of outliers and certain types of heteroscedasticity \citep{W84}. In addition, TPSMN distributions can also be used to capture unobserved heterogeneity that induces asymmetry of the residual errors \citep{RH15}. The implementation of these models is straightforward using the R package `twopiece' (available under request). We propose a general improper prior structure for the models of interest that covers certain priors obtained by formal rules. We show that the corresponding posterior is proper under mild conditions that can be extended to cases where the response variables are censored, a common phenomenon in survival analysis. The contribution of this paper consists mainly of extending the LRMs in \cite{RH15}, who only consider likelihood-based inference and prediction, to the Bayesian framework. The Bayesian approach provides natural tools, namely, the posterior predictive distribution, for conducting prediction about right-censored responses. This paper also presents a tractable alternative strategy to that proposed in \cite{RG16} to flexibly modelling the errors in LRMs. The rest of the paper is organised as follows. In Section \ref{TPDistributions}, we present the family of distributions of interest (TPSMN) and briefly discuss some of their properties. In Section \ref{LRMTP}, we describe the LRMs with TPSMN errors and the proposed prior structure, and then provide sufficient conditions for the propriety of the corresponding posterior distribution. In Section \ref{AFTModels}, we discuss the propriety of the posterior distribution in cases when the response variables are censored. We link these results with survival regression models. In Section \ref{SimulationStudy}, we present a simulation study which shows good frequentist performance of the proposed models. In Section \ref{Example} we present two examples with real data in the context of survival times of cancer patients. Proofs of the results as well as tables associated to the simulation study in Section \ref{SimulationStudy} are presented in the Supplemental Material.

\section{Background on two--piece distributions}\label{TPDistributions}

Let us first recall the definition of two-piece distributions. We refer the reader to \cite{RS14} and \cite{RH15} for a more extensive discussion on these models. A real random variable $Z$ is said to be distributed according to a two-piece distribution, denoted $Z\sim \mbox{TP}(\mu,\sigma,\delta,\gamma;f)$, if its probability density function (PDF) can be written as:
\begin{eqnarray}\label{TPPDF}
g(z\vert\mu,\sigma,\delta,\gamma) = \dfrac{2}{\sigma[a(\gamma)+b(\gamma)]}\left[f\left(\dfrac{ z-\mu }{\sigma b(\gamma)}\Big\vert\delta\right)I(z<\mu) + f\left(\dfrac{ z-\mu }{\sigma a(\gamma)}\Big\vert\delta\right)I(z\geq \mu)\right],\,\,\, z\in{\mathbb R},
\end{eqnarray}
where $f$ is a symmetric PDF with support on ${\mathbb R}$ and mode at $0$, $\mu\in{\mathbb R}$ is a location parameter and the mode of the density, $\sigma\in{\mathbb R}_+$ is a scale parameter, $\delta\in\Delta\subset{\mathbb R}$ is a shape parameter, $\gamma\in \Gamma\subset {\mathbb R}$ is a skewness parameter, and $\{a(\gamma),b(\gamma)\}$ are positive functions of the parameter $\gamma$. Several parameterisations $\{a(\cdot),b(\cdot)\}$ of these models are studied in \cite{A05} and \cite{RS14}. In our applications we will adopt the parameterisation proposed in \cite{MH00}: $\{a(\gamma),b(\gamma)\}=\{1-\gamma,1+\gamma\}$, $\gamma\in(-1,1)$. Some properties of this family of distributions are presented below.
\begin{enumerate}
\item The tail behaviour of (\ref{TPPDF}) is the same in each direction.
\item The moments of (\ref{TPPDF}) exist whenever the moments of the baseline PDF $f$ exist.
\item The Fisher information matrix associated to this sort of models is well defined \citep{RS14}, in contrast to some skew-symmetric models, such as the Azzalini's skew-normal distribution \citep{A85}.
\item Despite the fact that the PDF (\ref{TPPDF}) is not twice differentiable at the mode, \cite{A05} showed that the maximum likelihood estimators of the parameters of (\ref{TPPDF}) have good asymptotic properties.
\end{enumerate}
 Throughout, we focus on the case where $f$ belongs to the family of SMN.  Recall also that a symmetric PDF $f$ is said to be a SMN if it can be written as:
\begin{eqnarray}\label{SMN}
f(z\vert\delta) = \int_{{\mathbb R}_+} \tau^{1/2}\phi(\tau^{1/2} z) d H(\tau\vert \delta),
\end{eqnarray}
where $H$ is a mixing distribution with positive support, $\phi$ represents the standard normal PDF, and $\delta\in \Delta \subset{\mathbb R}$ is a shape parameter. This family contains distributions of great interest in practice such as the normal distribution, Logistic distribution, Laplace distribution, generalysed hyperbolic distribution, and the Student-$t$ distribution.

From the expressions in \cite{A05} we can obtain the cumulative distribution (CDF) associated to (\ref{TPPDF}) as follows:
\begin{eqnarray}\label{TPCDF}
G(z\vert\mu,\sigma,\delta,\gamma) &=& \dfrac{2 b(\gamma)}{a(\gamma)+b(\gamma)} F\left(\dfrac{z-\mu}{\sigma b(\gamma)}\Big\vert \delta\right)]I(z<\mu) \notag\\
&+& \dfrac{b(\gamma)-a(\gamma)}{a(\gamma)+b(\gamma)} + \dfrac{2 a(\gamma)}{a(\gamma)+b(\gamma)} F\left(\dfrac{z-\mu}{\sigma a(\gamma)}\Big\vert\delta\right) I(z\geq \mu),\,\,\, z\in{\mathbb R},
\end{eqnarray}
where $F$ is the CDF associated to the PDF $f$. From the latter expression we can see that ${\mathbb P}[Z \leq \mu]=P_{\gamma}=\frac{b(\gamma)}{a(\gamma)+b(\gamma)}$. That is, the parameter $\mu$ is the $P_{\gamma}-$th quantile of $Z$, and the parameter $\gamma$ controls the allocation of mass on either side of the mode $\mu$. The PDF, CDF, quantile function, and random number generation of two-piece distributions are implemented in the R package `twopiece', which is available under request. Some examples of the shape of the density \eqref{TPPDF}, for some choices of the baseline density $f$, are presented in the Supplemental Material.

\section{Linear regression with two-piece errors}\label{LRMTP}

Consider the linear regression model:
\begin{eqnarray}\label{LRM}
y_j= {\bf x}_j^{\top}\bm{\beta} + \varepsilon_j,
\end{eqnarray}
where $y_j\in{\mathbb R}$, $j=1,\dots,n$, $\bm{\beta}$ is a $p$-dimensional vector of regression parameters, $\varepsilon_j \stackrel{i.i.d.}{\sim} \mbox{TP}(0,\sigma,\delta,\gamma;f)$, $f$ is a SMN, and ${\bf X}=({\bf x}_1^{\top},\dots,{\bf x}_n^{\top})^{\top}$ is a known $n\times p$ design matrix of full column rank. The resulting model is centred at the mode of the distribution of the errors (which is $0$), which represents the $P_{\gamma}-$th quantile. We can re-centre the model at the mean, provided it exists, or any quantile of interest by adjusting the intercept after obtaining estimators for the corresponding parameters. The likelihood function associated to these assumptions is given by:
\begin{eqnarray}
s({\bf y}\vert \bm{\beta},\sigma,\delta,\gamma) = \prod_{j=1}^n  s(y_j - {\bf x}_j^{\top}\bm{\beta}\vert 0,\sigma,\delta,\gamma),
\end{eqnarray}
where $s$ is the PDF given by (\ref{TPPDF}). We adopt the prior structure:
\begin{eqnarray}\label{PriorStructure}
\pi({\bm \beta},\sigma,\delta,\gamma) \propto \dfrac{\pi(\gamma)\pi(\delta)}{\sigma^q},
\end{eqnarray}
\noindent where $q\geq 0$ and $\pi(\gamma)$ and $\pi(\delta)$ are proper priors. This prior structure covers the structure of some priors obtained by formal rules, for specific choices of the power hyperparameter $q$ and the priors $\pi(\gamma)$ and $\pi(\delta)$. For instance, for the choices $q=1$ and $\pi(\gamma)\propto(1-\gamma)^{-\frac{1}{2}}$, the prior \eqref{PriorStructure} corresponds to the independence Jeffreys prior (see \cite{RS14} for a study of this prior in the context of location-scale TPSMN). Expressions for the reference prior and the Jeffreys prior have not been calculated, but we conjecture that they have a similar structure to that of prior \eqref{PriorStructure}. Their calculation represents a possible research direction. The following result provides general conditions for the propriety of the posterior distribution under the prior \eqref{PriorStructure}.

\begin{theorem}\label{ProperLRM}
Consider the model (\ref{LRM})--(\ref{PriorStructure}), where $\varepsilon_j \stackrel{i.i.d.}{\sim} \mbox{TP}(0,\sigma,\delta,\gamma;f)$ and $f$ is a SMN. Consider the following conditions:

\begin{enumerate}[(i)]
\item The posterior associated to the linear regression model (\ref{LRM}), with errors distributed according to the symmetric baseline distribution $f$, together with the prior $\pi({\bm \beta},\sigma,\delta) \propto \sigma^{-q}\pi(\delta)$ is proper.
\item $\int_{\Gamma} \dfrac{h(\gamma)^{n+q-1}}{[a(\gamma)+b(\gamma)]^n} \pi(\gamma)  d\gamma <\infty$, where $h(\gamma)=\min\{a(\gamma),b(\gamma)\}$,
\item $\int_{\Gamma} \dfrac{H(\gamma)^{n+q-1}}{[a(\gamma)+b(\gamma)]^n} \pi(\gamma) d\gamma <\infty$, where $H(\gamma)=\max\{a(\gamma),b(\gamma)\}$.
\end{enumerate}
Then, (i) and (ii) are necessary conditions, while (i) and (iii) are sufficient conditions for the propriety of the posterior distribution of $({\bm \beta},\sigma,\delta,\gamma)$.
\end{theorem}

This result indicates that, in order to check the propriety of the posterior of $({\bm \beta},\sigma,\delta,\gamma)$, we only need to check the propriety of the posterior associated to the underlying model with residual errors distributed according to the symmetric baseline distribution $f$ together with a condition on the parameterisation  $\{a(\gamma),b(\gamma)\}$. In particular, for $q=1$, conditions (ii) and (iii) are satisfied by any choice of  $\{a(\gamma),b(\gamma)\}$. Moreover, if the functions $a(\cdot)$ and $b(\cdot)$ are bounded and $q >1$, conditions (ii) and (iii) are automatically satisfied. The parameterisation $\{a(\gamma),b(\gamma)\}=\{1-\gamma,1+\gamma\}$, proposed in \citep{MH00}, $\gamma \in (-1,1)$, satisfies this boundedness condition. For unbounded parameterisations and $q>1$ (such as the one proposed by \cite{FS98}: $\{a(\gamma),b(\gamma)\}=\{\gamma,1/\gamma\}$, $\gamma>0$), the finiteness condition in (iii) depends on the choice of the prior $\pi(\gamma)$.

The following result presents conditions for the existence of the posterior for the case when the baseline density $f$ belongs to the family of SMN and $q=1$.
\begin{corollary}\label{ProperSMN}
Consider the model (\ref{LRM})--(\ref{PriorStructure}) with $q=1$, where $\varepsilon_j \stackrel{i.i.d.}{\sim} \mbox{TP}(0,\sigma,\delta,\gamma;f)$ and $f$ is a SMN. Then, the posterior distribution of $({\bm \beta},\sigma,\gamma)$ is proper provided that ${\bf y} \not\in {\mathcal C}({\bf X})$, where ${\mathcal C}({\bf X})$ denotes the column space of ${\bf X}$, $n>p$, together with condition (iii) from Theorem \ref{ProperLRM}.
\end{corollary}
This result is satisfied with probability one since the distribution of the residual errors is continuous. For the case when $q>1$, conditions for the existence of the posterior associated to the model with symmetric errors become more restrictive. Next, we present some particular cases where the propriety of the posterior distribution can be easily checked.
\begin{corollary}\label{ProperNLLG}
Consider the model (\ref{LRM})--(\ref{PriorStructure}) and suppose that the baseline density $f$ in (\ref{TPPDF}) is either a normal distribution, a Logistic distribution, a Laplace distribution, or a generalised hyperbolic distribution with fixed shape parameter. Suppose that ${\bf y} \not\in {\mathcal C}({\bf X})$, $n>p+1-q$, and condition (iii) from Theorem (\ref{ProperLRM}) are satisfied. Then, the posterior distribution $({\bm \beta},\sigma,\gamma)$ is proper.
\end{corollary}
Model \eqref{LRM}--\eqref{PriorStructure} can be implemented by using the `twopiece' R package. Moreover, several Markov Chain Monte Carlo (MCMC) samplers have been developed for this kind of models. For instance, \cite{W14} propose a blocked Metropolis-within-Gibbs algorithm that takes advantage of the representation of SMN distributions. For the case when $f$ in (\ref{TPPDF}) is a Laplace distribution, \cite{W13} proposed an alternative MCMC algorithm based on a uniform mixture representation of the Laplace distribution.

We point out that Bayesian quantile LRMs \citep{YM01} represent a family of models closely related to \eqref{LRM}. These models can be interpreted as the LRM \eqref{LRM}, with $\epsilon_j$ distributed according to a two-piece Laplace distribution, where the parameter $\gamma$ is fixed according to the quantile of interest specified by the user. We emphasise that, in our context, we do not fix the parameter $\gamma$ but, instead, obtain posterior inference about this parameter using the prior structure \eqref{PriorStructure}.
\subsection*{Choice of the prior for $\gamma$}
Rubio and Steel \cite{RS14} proposed a prior elicitation strategy for the parameter $\gamma$, based on the interpretation of this parameter, that can be used to construct a weakly informative proper prior. They propose assigning a $\text{Beta}(a_0,b_0)$ prior on a measure of skewness which is an injective function of the parameter $\gamma$. This strategy induces a proper prior on $\gamma$ which can be used to construct informative and noninformative priors on $\gamma$:
\begin{eqnarray*}
\pi(\gamma) \propto \dfrac{\vert a^{\prime}(\gamma)b(\gamma) - a(\gamma)b^{\prime}(\gamma) \vert}{[a(\gamma)+b(\gamma)]^{a_0+b_0}}a(\gamma)^{a_0-1}b(\gamma)^{b_0-1},
\end{eqnarray*}
where $a^{\prime}(\cdot)$ and $b^{\prime}(\cdot)$ denote the derivatives of $a(\cdot)$ and $b(\cdot)$, respectively. For the case when $a_0=b_0=1/2$, coupled with the parameterisation in \cite{MH00}, this strategy leads to the Jeffreys prior of $\gamma$ \citep{RS14}. If $a_0=b_0=1$, this strategy leads to a uniform prior on $\gamma\in(-1,1)$. Throughout this paper we adopt this prior with $a_0=b_0=1/2$, this is $\pi(\gamma)\propto (1-\gamma^2)^{-\frac{1}{2}}$. This prior has been shown to induce a posterior distribution with good frequentist properties in the context of location-scale models \citep{RS14}.
%
%


\section{Accelerated failure time models}\label{AFTModels}
\subsection{Propriety results}
AFT models are of great interest in survival analysis given that they can be used for modelling a set of survival times ${\bf T}=(T_1,\dots,T_n)$ in terms of a set of covariates $\bm{\beta}$ through the model equation:
\begin{eqnarray}\label{AFT}
y_j =\log(T_j) = {\bf x}_j^{\top}\bm{\beta} + \varepsilon_j, \,\,\, j=1,\dots,n,
\end{eqnarray}
where $\bm{\beta}$ is a $p$-dimensional vector of regression parameters, and ${\bf X}=({\bf x}_1^{\top},\dots,{\bf x}_n^{\top})^{\top}$ is a known $n\times p$ design matrix of full column rank. The use of normal and Logistic residual errors represent the most common distributional assumptions. Other distributional assumptions were discussed Section \ref{intro} and in \cite{RH15}.

We assume that $\varepsilon_j \stackrel{i.i.d.}{\sim} \mbox{TP}(0,\sigma,\delta,\gamma;f)$, where the baseline density $f$ is a SMN. If we adopt the prior structure (\ref{PriorStructure}) for this model, then the corresponding posterior is proper under the conditions in Corollaries \ref{ProperSMN} and \ref{ProperNLLG}. However, a common challenge that arises in the context of the analysis of time-to-event data is the presence of censored observations (see \cite{RH15} for a discussion on this). The following result provides sufficient conditions for the propriety of the posterior distribution for the case when the sample contains both censored and uncensored observations.
\begin{theorem}\label{CensNonCens}
Consider the linear regresion model (\ref{AFT}) with prior (\ref{PriorStructure}). Suppose that $n_c\leq n$ survival times are censored and $n_o=n-n_c$ are observed. Let ${\bf y}_o$ be the set of uncensored observations and ${\bf X}_o$ be the corresponding design matrix. Then, the posterior distribution of $(\bm{\beta},\sigma,\delta,\gamma)$ is proper provided that the posterior distribution associated to the $n_o$ uncensored observations is proper.
\end{theorem}
Since this result relies only on the sub-sample of uncensored observations, we can use the results in the previous section to check the propriety of the posterior. Corollary \ref{ProperSMN} provides conditions for the case when $f$ is a SMN and $q=1$, while Corollary \ref{ProperNLLG} provides conditions for the case when $q\geq1$ and certain particular choices of $f$.

An extreme case that arises in practice is when the sample contains only censored observations. The next result presents sufficient conditions for the existence of the posterior distribution in this scenario.
\begin{corollary}\label{OnlyCensSMN}
Consider the model (\ref{AFT}) with prior (\ref{PriorStructure}). Suppose that $f$ is a scale mixture of normals, $q=1$, $n_I\leq n$ observations are interval censored, where the length of these intervals is finite, and that the other $n-n_I$ observations are censored of any other type. Denote the $n_I$ interval-censored observations as $(I_1,\dots,I_{n_I})$, and let ${\bf X}_{n_I}$ be the corresponding design submatrix. Then, the corresponding posterior is proper if ${\mathcal E} = I_1\times \dots \times I_{n_I}$ and the column space of ${\bf X}_{n_I}$ are disjoint, together with the condition $n_I > p $, and condition (iii) from Theorem \ref{ProperLRM}.
\end{corollary}
Similarly, for $q>1$ we have the following results.
\begin{corollary}\label{OnlyCensNLLG}
Consider the model (\ref{AFT}) with prior (\ref{PriorStructure}). Suppose that $f$ is either a normal, Logistic, Laplace or generalised hyperbolic distribution; $n_I\leq n$ observations are interval censored, where the length of these intervals is finite, and that the other $n-n_I$ observations are censored of any other type. Then, the corresponding posterior is proper if ${\mathcal E} = I_1\times \dots \times I_{n_I}$ and the column space of ${\bf X}_{n_I}$ are disjoint, together with the condition $n_I > p + 1 - q $, and condition (iii) from Theorem \ref{ProperLRM}.
\end{corollary}
As discussed in \cite{RG16}, checking that ${\mathcal E}$ and the column space of ${\bf X}_{n_I}$ are disjoint can be formulated as a linear programming problem (LP). Denote $\bm{\eta}\in {\mathbb R}^{p}$, $\bm{\xi}=(\xi_1,\dots,\xi_{n_I})\in {\mathcal E}$, and $I_j = [l_j,u_j]$, $j=1,\dots,n_I$. Define the LP problem:
\begin{eqnarray}\label{LPP}
\text{Find}&& \max_{\bm{\eta},\bm{\xi}} 1,\notag\\
\text{Subject to}&& {\bf X}_{n_I}\bm{\eta} = \bm{\xi},\notag\\
\text{and}&&\log(l_j)\leq \xi_j \leq \log(u_j),\,\,\, j=1,\dots,n_I.
\end{eqnarray}
Thus, the disjointness condition is equivalent to verifying the infeasibility of the LP problem (\ref{LPP}), for which there are several theoretical and numerical tools (LP solvers) available \citep{DT97}. It is \emph{important} to notice that the optimisation step in (\ref{LPP}) represents just a tool to connect the propriety conditions in Corollaries \ref{OnlyCensSMN} and \ref{OnlyCensNLLG} with the feasibility of the restrictions in a LP problem.

\section{Simulation study}\label{SimulationStudy}

In this section we present a simulation study that illustrates the performance of the proposed prior structure. We adopt the simulation scenarios used in \cite{RH15} in order to allow for qualitative comparisons. We study the LRM:
\begin{eqnarray}\label{LRMSim}
y_j = \beta_1 + \beta_2x_{1j} + \beta_3x_{2j} + \varepsilon_j, \,\,\, j=1,\dots,n,
\end{eqnarray}
where we simulate the variables $x_{1j}$ and $x_{2j}$ from a standard normal distribution and consider different combinations of the distribution of the residual errors and the sample size $n$.

In the first scenario, we simulate the residual errors from a two-piece normal distribution with unit scale parameter and skewness parameter $\gamma=0,0.25,0.5,0.75$, $(\beta_1,\beta_2,\beta_3)=(1,2,3)$, and $n=100,250,500$. We fit the LRM (\ref{LRMSim}) with $\varepsilon_j\stackrel{i.i.d.}{\sim} \mbox{TP}(0,\sigma,\gamma;f)$, where $f$ is the standard normal PDF. We adopt the product prior structure (\ref{PriorStructure}) with $q=1$ and the Jeffreys prior on the parameter $\gamma$. For each of these scenarios, we obtain $N=1,000$ samples of size $2,000$ from the posterior distribution using the R t-walk sampler \citep{CF10} after a burn-in period of $5,000$ iterations and thinned to every $25$th iteration (this is, a chain of length $55,000$). Then, we calculate the proportion of $95\%$ credible intervals that include the true value of the parameter, the median posterior estimators, the maximum \emph{a posteriori} (MAP) estimators, the maximum likelihood estimators (MLEs) for comparison, as well as the median of the Bayes factors associated to the hypothesis $H_0:\gamma=0$ (approximated using the Savage-Dickey density ratio).

In the second and third scenarios, we simulate the residual errors from a two-piece Student-$t$ distribution with degrees of freedom $\delta=2,5$. For these scenarios we fit the LRM (\ref{LRMSim}) with $\varepsilon_j\stackrel{i.i.d.}{\sim} \mbox{TP}(0,\sigma,\delta,\gamma;f)$, where $f$ is the Student-$t$ PDF with $\delta>0$ degrees of freedom. We adopt the prior structure (\ref{PriorStructure}) with $q=1$ and the Jeffreys prior on the parameter $\gamma$. For the degrees of freedom $\delta$, we use the approximation to the Jeffreys prior for this parameter proposed in \cite{JS10}:
\begin{eqnarray}\label{PriorDFST}
\pi(\delta) = \dfrac{2d\delta}{(\delta+d)^3}.
\end{eqnarray}
We choose the hyperparameter $d=10$, which induces a prior with mode at $\delta=5$. In the fourth scenario, we simulate from the linear regression model:
\begin{eqnarray*}
\log(y_j) = {\bf x}_j^{\top}\bm{\beta} + \varepsilon_j, \,\,\, j=1,\dots,n,
\end{eqnarray*}
with $n=100,250,500$, $\beta = (1,2,3)^{\top}$, and ${\bf x}_j = (1,x_{j1},x_{j2})^{\top}$. The second and third entries of the covariates ${\bf x}_j$ are simulated from a right-half-normal with scale parameter $1/3$. The errors $\varepsilon_j$ are simulated from a two-piece normal distribution with parameters $\mu=0$, $\sigma=0.25$, and $\gamma=0,0.25,0.5,0.75$. We truncate the observations $y_j$ that are greater than $17.5$, producing samples with 15\%--35\% censored observations. Results are reported in Tables \ref{table:SN100}--\ref{table:CSN500} of the Supplemental Material. In the first scenario we can observe a good coverage as well as good frequentist properties of the estimators associated to the proposed model overall. We can also observe that the Bayes factors clearly identify the case when the errors are symmetric. In the second and third scenarios we observe a good coverage of the credible intervals associated to the regression parameters $(\beta_1,\beta_2,\beta_3)$ as well as an accurate point estimation. However, in order to get a decent coverage of the credible intervals associated to the scale an tail parameters $(\sigma,\delta)$, we need at least 250 observations. This is a well known phenomenon about the estimation of degrees of freedom of the Student-$t$ distribution. Interestingly, the level of skewness does not seem to affect the performance of the credible intervals, even though for the case when $\gamma=0$ we are fitting an overparameterised model. Although the proposed model performs well even when the true distribution of the residual errors is symmetric, in practice, we recommend conducting a formal model selection between the models with symmetric and asymmetric errors in order to avoid overparameterisation, which has a other unpleasant effects such as increasing the length of the credible intervals. The presence of mild levels of censored observations does not greatly affect the performance of the Bayes estimators and the coverage proportions as we can see from Tables \ref{table:CSN100}--\ref{table:CSN500} in the Supplemental Material.

\pagebreak

\section{Applications}\label{Example}

In this section we present two examples with publicly available real data to illustrate the usefulness and performance of the proposed Bayesian LRMs. Both examples concern the study of the survival times of cancer patients. In the second example, we discuss the impact of using flexible errors in terms of prediction. For the models with two-piece residual errors, we employ the parameterisation in \cite{MH00}. Posterior samples are obtained using the R twalk sampler \citep{CF10}. Since the implementation of the log-posterior associated to the models of interest is very tractable, other samplers (such as Metropolis-Hastings or Metropolis-within-Gibbs samplers) can also be easily implemented. R codes and data used for these examples are available upon request.

Model comparison is conducted in terms of three formal model selection tools: Bayesian information criterion (BIC), Bayes factors, and log-predictive marginal likelihood (LPML, \cite{S99}). Bayes factors are calculated using an importance sampling technique. The use of the Bayes factors with the proposed improper prior structure is justified since we are employing improper priors only on the common parameters of the different models, while the priors on the shape parameters have the same interpretation across the different models (see \cite{VS15} for a discussion on this point). Bayes factors and BIC are useful to identify the model that provides the best fit. On the other hand, LPML is a measure that ranks the models of interest in terms of their predictive performance \citep{S99}. Therefore, these two variables provide complementary information. Their combination is particularly relevant in survival analysis given that we are interested on selecting the best model for the data but, since this model is often used for prediction of the residual life of patients that survived beyond the end of the study (see \cite{RH15}), it is important to check that the model also has a better predictive performance than the competitor models.

\subsection{Small Cell Cancer Data}

We analyse the data set from \cite{Y95} about a lung cancer study with two different types of treatment. The data set contains $n=121$ survival times (in days) of patients with small cell lung cancer (SCLC) that were administrated two types of therapies. For patients with SCLC the standard treatment consists of a combination of etoposide (E) and cisplatin (P); however the optimal order for the administration of these two treatments has not been established \citep{Y95}. The group of patients was splitted into two groups: Arm A (62 patients), whose therapy consisted of P followed by E, and Arm B (59 patients), whose therapy consisted of E followed by P. The covariates used for this study are the ``Entry age'' (in years) and the type of treatment (Arm A and Arm B). The sample contains $n_c=23$ right-censored observations. We fit an AFT model (\ref{AFT}) with 4 distributional assumptions for the residual errors: two-piece Laplace (TP Laplace), two-piece Normal (TP Normal), as well as corresponding symmetric submodels (Laplace and Normal). The propriety of the corresponding posterior distributions is guaranteed by Theorem \ref{ProperSMN} and Corollary \ref{CensNonCens}. We adopt the prior structure (\ref{PriorStructure}) with $q=1$ and the Jeffreys prior on the skewness parameter $\gamma\in(-1,1)$. For each of these models, a sample of size $10,000$ was obtained from the posterior distribution after a burn-in period of $50,000$ iterations and thinned to every $25$ iterations (this is, $300,000$ MCMC iterations in total). Table \ref{table:SCLC} presents a summary of the posterior samples as well as the model comparison tools. The TP Laplace model performs better overall (closely followed by the TP Normal model) in terms of BIC, LPML and Bayes factors, which suggests the need for a model with heavier tails than normal and asymmetry.

\begin{table}[h!]
\begin{center}
\scriptsize
\begin{tabular}[h]{|c|c|c|c|c|}
\hline
Model &  TP Laplace & TP Normal & Laplace & Normal \\
\hline
Intercept  &  6.690 (5.964,7.445) & 7.150   (6.282,8.150 )   &  7.114 (6.168,8.294)  & 7.633 (6.560,8.653)   \\
Entry age   &  -0.009 (-0.021,0.004) & -0.016   (-0.031,-0.002)    & -0.011 (-0.029,0.004)  &   -0.017 (0.033, -0.0003)  \\
Treatment  & -0.446 (-0.682,-0.197) &  -0.387  (-0.660,-0.107)    & -0.403 (-0.683,-0.137)  & -0.408 (-0.703,-0.139)   \\
$\sigma$  & 0.650 (0.517,0.792) &  0.785  (0.664,0.915)    & 0.648 (0.532,0.795)  & 0.759 (0.660,0.890)   \\
$\gamma$  & -0.395 (-0.599,-0.1872 ) &  -0.383  (-0.639,-0.116)   &  -- &  --  \\
BIC & {\bf 283.09} &  284.75 &  292.49 &   287.59  \\
Bayes factor &  -- &  0.721  & 0.016  &  0.174  \\
LPML & {\bf -134.893} & -136.416 & -141.0203 & -138.1511\\
\hline
\end{tabular}
\caption{\small SCLC Lung Cancer data: Posterior median and 95\% credible intervals and model comparison tools. The Bayes factors are calculated against the model with TP Laplace errors.}
\label{table:SCLC}
\end{center}
\end{table}

\pagebreak
\subsection{North Central Cancer Treatment Group (NCCTG) Lung Cancer Data}\label{Lung}

In this application we analyse the NCCTG Lung Cancer data set, which is available in the `survival' R package. The data set with complete cases (removing missing covariates) contains the survival times (in days) of $n=227$ patients with advanced lung cancer from the NCCTG. The sample contains $n_c=63$ right-censored observations. The aim of this study was to compare the information from a questionnaire applied to a group of patients against the information obtained by the patient's physician in terms of prognostic power \citep{L94}. We fit an AFT model with three covariates (``age'' (in years),``sex'' (Male=1 Female=2), ``ph.ecog'' [ECOG performance score, 0=good--5=dead]) as well as an intercept with 4 residual error distributions: two--piece Logistic errors (TP Logistic), two-piece normal errors (TP Normal), and the corresponding symmetric sub-models (Logistic and normal). We adopt the prior structure (\ref{PriorStructure}) with $q=1$ and the Jeffreys prior on $\gamma$. The propriety of the corresponding posterior distributions is guaranteed by Theorem \ref{ProperSMN} and Corollary \ref{CensNonCens}. We obtain a posterior sample of size $10,000$ after a burn-in of $50,000$ and  thinned to every $25$th iteration ($300,000$ MCMC iterations in total). Table \ref{table:NCCTG} shows a summary of the posterior samples as well as the model comparison tools. The model with TP Logistic errors performs better overall, which suggests the presence of skewness and slightly heavier tails than normal.

%

\begin{table}[h!]
\begin{center}
\scriptsize
\begin{tabular}[h]{|c|c|c|c|c|}
\hline
Model &  TP Logistic & TP Normal & Logistic & Normal \\
\hline
Intercept  & 6.531 ( 5.514, 7.565) &  6.940  (5.840,7.979) & 5.965 ( 4.985, 6.962) & 6.477 (5.309, 7.628)  \\
Age     & -0.010 (-0.025, 0.004) &  -0.015 ( -0.029, 0.001) &  -0.008   (-0.023,  0.006)  &  -0.018 (-0.034, -0.002) \\
Sex      & 0.435 (0.188, 0.720) & 0.446  (0.197,0.726) &  0.496   (0.222, 0.761)  & 0.529  (0.231, 0.842) \\
ph.ecog  & -0.363 (-0.533, -0.167) &  -0.326 (0.507,-0.119) &   -0.407   (-0.601, -0.221)  &  -0.359 (-0.571,-0.157)\\
$\sigma$ & 0.495 (0.429, 0.569) & 0.906  (0.806,1.018) &  0.548  (0.479,0.628)  &  1.043 (0.929, 1.170)\\
$\gamma$ &  0.384 (0.129, 0.600) & 0.481  (0.270,0.669) &   --  & -- \\
BIC & {\bf 556.60} & 566.50  & 562.17 & 580.96 \\
Bayes factor & -- & 0.006 & 0.019 &  2$\times 10^{-6}$ \\
LPML & {\bf -268.966} & -274.415  &  -272.91 & -283.23\\
\hline
\end{tabular}
\caption{\small NCCTG Lung Cancer data: Posterior median and 95\% credible intervals and model comparison tools. The Bayes factors are calculated against the model with TP Logistic errors.}
\label{table:NCCTG}
\end{center}
\end{table}

We now analyse the impact of using more flexible errors in terms of prediction. As discussed in \cite{RH15} and \cite{RG16}, it is often of interest to study the distribution of the residual life of patients that survived beyond the end of the study. In order to obtain these predictions, consider the AFT model (\ref{AFT}) with a general residual error distribution $\varepsilon_j \stackrel{i.i.d.}{\sim} \mbox{TP}(c(\sigma,\gamma),\sigma,\gamma;f)$, where $c(\sigma,\gamma)$ denotes the point at which the AFT model is centred (\emph{e.g.}~the mode, mean, or median). Denote by $\bm{\theta}=(\bm{\beta},\sigma,\gamma)$ the model parameters, and let $\pi(\bm{\theta})$ be the corresponding prior. Suppose that the $j$th subject survived beyond time $T_j$, and therefore the corresponding observation is right-censored. Then, the posterior predictive CDF of the residual life for this subject is given by:
\begin{eqnarray}\label{PredResCDF}
\Pi_R(t\vert {\bf T},j) = \dfrac{\Pi(t\vert {\bf T},j)-\Pi(T_j\vert {\bf T},j)}{1-\Pi(T_j\vert {\bf T},j)},\,\,\, t>T_j,
\end{eqnarray}
where
\begin{eqnarray*}
\Pi(t\vert {\bf T},j) = \int_0^t \pi(r\vert {\bf T},j) dr,
\end{eqnarray*}
is the posterior predictive CDF associated to this model, and
\begin{eqnarray}\label{PredDist}
\pi(r\vert {\bf T},j) = \int \dfrac{1}{r}s[\log r\vert \bm{\theta},{\bf x}_j]\pi(\bm{\theta}\vert {\bf T}) d\bm{\theta},\,\,\, r>0,
\end{eqnarray}
is the posterior predictive PDF associated to subject $j$, and $\pi(\bm{\theta}\vert {\bf T})$ represents the posterior distribution of $\bm{\theta}$. We recommend centring the model (after sampling from the posterior distribution) around the median rather than the mean since the latter may not exist for certain combinations of the distribution of the residual errors and priors. Moreover, median estimators are robust to the presence of outliers. The posterior predictive survival function of the residual life of subject $j$ is given by ${\mathcal S}_R(t\vert {\bf T},j)=1-\Pi_R(t\vert {\bf T},j)$. This estimator takes into account the uncertainty on the model parameters given that they are integrated out with respect to the corresponding posterior distribution in (\ref{PredDist}). If we have a sample from the posterior distribution $\pi(\bm{\theta}\vert {\bf T})$, then we can approximate (\ref{PredResCDF}) by using a Monte Carlo approximation of (\ref{PredDist}). One advantage of the predictive estimator (\ref{PredResCDF}) over the plug-in estimator proposed in \cite{RH15} is that this incorporates the posterior uncertainty about the model parameters.

Table \ref{table:NCCTGPred} presents a summary of the quantiles of the residual life distributions, for the first 5 censored patients, using the AFT models with TP Logistic and Logistic errors centred at the median. As we discussed before, the model with TP Logistic errors produces a better fit and a better predictive performance. The quantiles higher than 50\% associated to the model with Logistic errors are much larger than those obtained with the model with TP Logistic errors. Therefore, although the inference on the regression parameters is very similar for these two models (see Table \ref{table:NCCTG}), the corresponding prediction intervals are very different. These differences can be explained using the expressions \eqref{PredResCDF}--\eqref{PredDist} which indicate the dependence of the predictions on the residual error distribution and the posterior distribution (which are different in this case).

\begin{table}[h!]
\begin{center}
\scriptsize
\begin{tabular}[h]{|c|c|c|c|c|c|}
\hline
Quantile & 5\% &  25\% & 50\% & 75\% & 95\% \\
\hline
TP Logistic model &&&&&\\
Patient 1  & 1037.6 & 1163.9 & 1382.0 & 1774.0 & 2976.8 \\
Patient 2  & 1040.2 & 1126.7 & 1287.3 & 1605.0 & 2641.4 \\
Patient 3  & 992.4  & 1117.3 & 1332.4 & 1719.3 & 2886.6 \\
Patient 4  & 868.8  & 1078.9 & 1406.8 & 1927.5 & 3358.7 \\
Patient 5  & 866.5  & 986.3  & 1187.5 & 1545.6 & 2604.5 \\
Logistic model &&&&&\\
Patient 1  & 1043.9 &  1211.7 & 1549.4 & 2318.6 & 5700.5 \\
Patient 2  & 1052.9 & 1206.9 & 1520.3 & 2241.5 & 5453.9 \\
Patient 3  & 997.8 & 1160.0 & 1485.5 & 2225.7 & 5481.2 \\
Patient 4  & 857.6 & 1034.8 & 1378.8 & 2136.7 & 5387.6 \\
Patient 5  & 869.3 & 1013.5 & 1302.4 & 1956.2 & 4829.0 \\
\hline
\end{tabular}
\caption{\small NCCTG Lung Cancer data: Quantiles of the predictive residual life distribution for the Median TP Logistic and Logistic models.}
\label{table:NCCTGPred}
\end{center}
\end{table}

%

%

\section{Discussion}

We introduced a flexible class of LRM that can account for departures from normality of the residual errors in terms of heavy tails, asymmetry and certain kinds of heteroscedasticity. We proposed a general noninformative prior structure and provided easy to check conditions for the propriety of the corresponding posterior distribution. A simulation study suggests a good frequentist performance of the proposed Bayesian models. The propriety results cover cases when the response variables are censored, which allows for implementing the proposed Bayesian models in survival analysis. The implementation of these models is tractable using already available R packages. For instance, the R package `twopiece' provides commands for the implementation of the PDF and CDF associated to TPSMN distributions.

In the real data applications, we have compared the proposed models against appropriate competitors using different sorts of model comparison tools. In the context of survival analysis, we advocated for the use of model selection tools that provide information about the model that better fits the data, as well as tools that provide information about the predictive performance of the models. This is particularly important in cases when the selected model is used for predicting the remaining life of a censored subject.



\end{document}